\documentclass[conference]{IEEEtran}
\IEEEoverridecommandlockouts

\usepackage{cite}
\usepackage{amsmath,amssymb,amsfonts}
\usepackage{algorithmic}
\usepackage{graphicx}
\usepackage{textcomp}
\usepackage{xcolor}
\usepackage{tikz}
\usetikzlibrary{arrows.meta,positioning,fit,backgrounds,calc}
\usepackage{tabularx,booktabs}
\def\BibTeX{{\rm B\kern-.05em{\sc i\kern-.025em b}\kern-.08em
    T\kern-.1667em\lower.7ex\hbox{E}\kern-.125emX}}
\begin{document}

\title{Auditory Intelligence: \\ Understanding the World Through Sound}

\author{\IEEEauthorblockN{Hyeonuk Nam\thanks{This position paper consolidates research ideas I intend to pursue while transitioning from academia to industry. I welcome collaborators and partners interested in developing datasets, evaluation protocols, and lightweight pilots for ASPIRE, SODA, AUX, and AUGMENT}}
\IEEEauthorblockA{Korea Advanced Institute of Science and Technology,
South Korea \\
frednam@kaist.ac.kr}
}

\maketitle

\begin{abstract}
Recent progress in auditory intelligence has yielded high-performing systems for sound event detection (SED), acoustic scene classification (ASC), automated audio captioning (AAC), and audio question answering (AQA). Yet these tasks remain largely constrained to surface-level recognition—capturing what happened but not why, what it implies, or how it unfolds in context. I propose a conceptual reframing of auditory intelligence as a layered, situated process that encompasses perception, reasoning, and interaction. To instantiate this view, I introduce four cognitively inspired task paradigms—ASPIRE, SODA, AUX, and AUGMENT—those structure auditory understanding across time-frequency pattern captioning, hierarchical event/scene description, causal explanation, and goal-driven interpretation, respectively. Together, these paradigms provide a roadmap toward more generalizable, explainable, and human-aligned auditory intelligence, and are intended to catalyze a broader discussion of what it means for machines to understand sound.
\end{abstract}


\section{Introduction}
Large language models (LLMs) have significantly augmented human capabilities by automating repetitive and tedious tasks \cite{GPT4,llama2,deepseek}. With simple text or voice prompts, they can generate ideas, assist with research, create visual content, and engage in human-like conversation. Multimodal extensions now enable interaction not only via text but also through images and live video, leveraging smartphone cameras and displays to interpret visual input. Yet despite rapid progress in visual–linguistic understanding, their interaction with sound remains largely confined to speech: current systems can recognize and synthesize speech, but they do not reliably comprehend non-speech audio such as music, ambient soundscapes, or everyday acoustic events.

Imagine always-on LLMs that can interpret the ambient soundscape. They could detect hazardous cues (e.g., explosions, fire alarms), monitor health by tracking coughs and analyzing snoring during sleep, and infer situational context—whether a user is working, resting, or socializing—to decide whether to deliver or defer notifications. Just as human assistants rely on auditory context to coordinate and intervene, AI assistants and robots must understand non-speech sound to collaborate effectively in real-world environments.

Recent advances in auditory artificial intelligence have driven substantial progress across a wide range of tasks. Sound event detection (SED) has become a foundational capability, enabling applications such as AI-driven perception, smart environments, and bioacoustic monitoring \cite{CASSE, DCASEtask4, crnn, sedmetrics, PSDS, freqdeptalsp, jitter}. Beyond SED, extensive work spans automatic speech recognition (ASR) and speaker recognition/verification \cite{specaug, conformer, mpc, wav2vec2.0, hubert, ASP, SAP, tdyaccess, freqse, c2datt}, sound event recognition \cite{PANN, coughcam, etri, AST, beats}, and sound event localization and detection (SELD) \cite{seld2019, starss22, 2022t3report, Biseld}. Emerging areas such as automated audio captioning (AAC) \cite{dcaseaac, clotho, chatgptaugaac}, audio question answering, few-shot bioacoustic detection \cite{dcasebed2024, bioacousticstrf}, and computational modeling of human auditory perception \cite{prtfnet, brainstem, contHRTF} further expand the scope of auditory intelligence. In parallel, text- or label-conditioned generative models for sound synthesis have gained traction \cite{audioldm, audiogen, vifs}, opening new directions for sound representation learning and multimodal integration.

While these advances have strengthened the foundations of machine listening, many current approaches remain constrained to surface-level recognition—identifying what occurred without understanding why it occurred, what it implies, or how it relates to broader perceptual and social contexts \cite{CASSE}. These limitations indicate that our conception of auditory intelligence is still underspecified and motivate a shift toward more cognitively grounded, context-aware frameworks.

\begin{figure*}[th]
\centering
\begin{tikzpicture}[
  font=\small,
  >=Latex,
  node distance=10mm and 18mm,
  box/.style={draw, rounded corners=2pt, align=center,
              inner sep=4pt, minimum height=8mm, text width=42mm},
  widebox/.style={draw, rounded corners=2pt, align=center,
              inner sep=4pt, minimum height=10mm, text width=120mm}
]
\node[box] (input) {Auditory input\\(waveform, mel spectrogram)};
\node[box, below=of input] (aspire) {ASPIRE\\Acoustic structure-based parsing\\(spectro-temporal descriptors)};

\node[box, below left=of aspire, xshift=-6mm] (soda) {SODA\\Event $\to$ Context $\to$ Scene\\(structured description)};
\node[box, below=of aspire] (aux) {AUX\\Causal / explanatory reasoning};
\node[box, below right=of aspire, xshift=6mm] (augment) {AUGMENT\\Trigger, Event, Motivation, Goal};

\node[widebox, below=27mm of aux] (response)
{Agent response and interaction\\(speech, music, sound effects, actions)};
\node[box, below right=of aux, text width=48mm] (mm)
{Multimodal grounding\\(audio--text--image)};

\draw[-{Latex}] (input) -- (aspire);
\draw[-{Latex}] (aspire.south west) -- (soda.north east);
\draw[-{Latex}] (aspire) -- (aux);
\draw[-{Latex}] (aspire.south east) -- (augment.north west);

\draw[-{Latex}] (soda) -- (response);
\draw[-{Latex}] (aux) -- (response);
\draw[-{Latex}] (augment) -- (response);

\draw[-{Latex}] (aux.east) -- (mm.west);
\draw[-{Latex}] (soda.east) to[out=0,in=180] (mm.west);
\draw[-{Latex}] (augment.east) to[out=0,in=180] (mm.west);

\begin{scope}[on background layer]
  \node[draw, dashed, rounded corners=2pt, fit=(aspire), inner sep=8pt] (g1) {};
  \node[draw, dashed, rounded corners=2pt, fit=(soda)(aux)(augment), inner sep=10pt] (g2) {};
  \node[draw, dashed, rounded corners=2pt, fit=(response), inner sep=10pt] (g3) {};
\end{scope}

\node[anchor=north west, fill=white, inner sep=-1pt]
      at ($(g1.north west)+(2pt,-2pt)$) {  layer 1: perceptual recognition};
\node[anchor=north west, fill=white, inner sep=0pt]
      at ($(g2.north west)+(2pt,-2pt)$) {  layer 2: contextual reasoning};
\node[anchor=north west, fill=white, inner sep=0pt]
      at ($(g3.north west)+(2pt,-2pt)$) {  layer 3: generative interaction};
\end{tikzpicture}
\caption{Overview of the proposed paradigms. ASPIRE provides spectro-temporal evidence; SODA, AUX, and AUGMENT perform structured description, explanation, and intent inference; outputs feed agent responses and multimodal grounding. Note that SODA, AUX, and AUGMENT can also be developed without ASPIRE.}
\label{fig:overview}
\end{figure*}
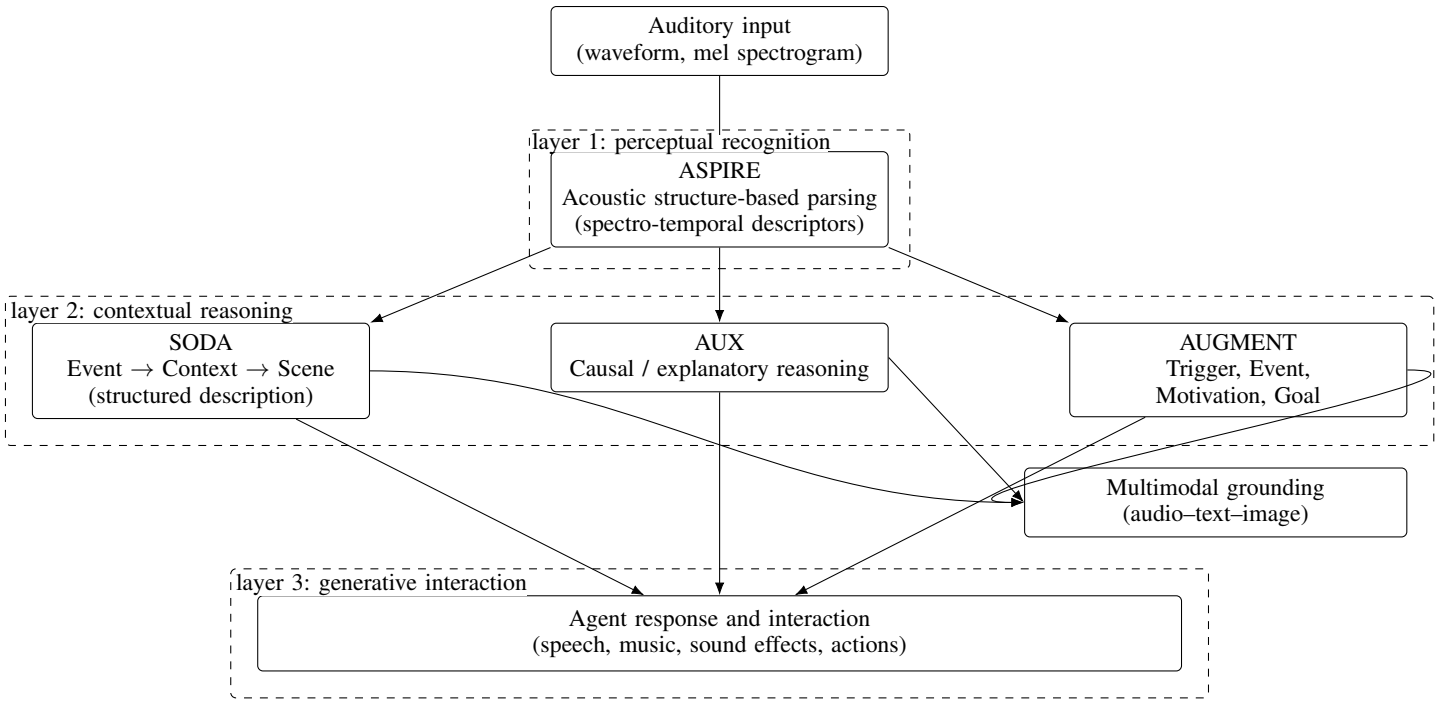

This paper reframes auditory intelligence as a layered, situated form of understanding rather than a set of task-specific recognition capabilities. I argue that sound, as perceived by intelligent agents, should not be treated as a raw signal to classify but as a cognitive medium encoding events, intentions, and contexts. From this perspective, auditory intelligence must evolve beyond pattern recognition to encompass causal reasoning, explanation, prediction, and multimodal integration. To operationalize this view, I propose four cognitively grounded task formulations and an accompanying conceptual framework that align machine hearing more closely with human auditory cognition. This work does not introduce a new model; instead, it offers a research lens and organizing scaffold for future datasets, benchmarks, and systems.

While the proposed paradigms build on existing tasks, they aim to extend auditory intelligence toward a more holistic and cognitively grounded direction. I anticipate a transitional period in which current benchmarks and the proposed formulations coexist, gradually converging toward richer, more human-aligned machine hearing. Although this work presents no quantitative evaluation, it is intended as a conceptual position and research agenda for the audio-AI community—guiding the development of acoustic models and audio–language models (ALMs) that enable AGI systems and AI agents to interact with humans and the world through sound in reliable, transparent, and effective ways.

\section{Auditory Intelligence}
Auditory intelligence denotes a machine’s ability to perceive, interpret, and act on sound in ways aligned with human auditory cognition. At its most basic, it entails recognizing acoustic events and mapping them to meaningful categories. Yet true auditory intelligence must extend beyond recognition to support causal inference and explanation, contextual reasoning, prediction, and context-sensitive action. Humans do not merely hear; they listen in context, infer intent, detect anomalies, anticipate consequences, and exhibit appropriate affective responses. Consequently, sound conveys not only what is happening, but also why it happens, where it occurs, and how it is likely to evolve.

Among the human senses, hearing plays a uniquely complementary and integrative role. Unlike vision, which offers high spatial resolution but a limited field of view, hearing affords omnidirectional coverage, temporal continuity, and access to events beyond line of sight \cite{CASSE}. It enables inferences about environmental structure, early detection of hazards and opportunities, and perception of others’ affect and intent—even in the absence of visual cues. Sound also mediates communication, coordination, and empathy through speech, music, and ambient signaling. For machines to interact intelligently and empathetically with people, they must treat sound not as raw data but as a cognitive medium embedded in physical and social context.

To make this notion precise, I conceptualize auditory intelligence as three interrelated cognitive layers:
\begin{enumerate}
    \item Perceptual recognition:\; identifying and categorizing acoustic patterns (e.g., speech, music, environmental events).
    \item Contextual reasoning:\; inferring causes, intent, affect, and the surrounding spatial/social context, including likely evolution of the scene.
    \item Generative interaction:\; producing context-appropriate linguistic or behavioral responses (e.g., explanations, dialogue, or actions).
\end{enumerate}

This layered model mirrors how humans use sound for adaptive perception and interaction. It also offers a scaffold for designing machine hearing systems that move beyond task-specific pattern recognition toward generalizable, explainable, and socially relevant intelligence. In the next section, I review how current auditory intelligence research addresses these cognitive functions, and where it still falls short.

\section{Related Works}
In recent years, auditory intelligence research has advanced across several core tasks. Sound event detection (SED) identifies and localizes acoustic events in continuous audio, typically using spectrogram-based neural networks or attention-equipped temporal models. Acoustic scene classification (ASC) assigns whole audio segments to predefined environmental categories (e.g., “office,” “street”), emphasizing holistic background perception. Automated audio captioning (AAC) maps audio to natural-language descriptions by training on paired audio–text datasets, commonly with encoder–decoder or sequence-to-sequence architectures.

Despite their technical success, these task formulations share a common limitation: they center on recognizing what is heard rather than understanding why it occurs, what it implies, or how it operates in context. SED and ASC often reduce sound to isolated labels or events, overlooking causal relationships and spatial or social dynamics. Although AAC provides richer outputs, most current models produce surface-level captions with limited reasoning, inference, or affective nuance. Moreover, these tasks are typically developed and evaluated in isolation, lacking a unified framework that links recognition, understanding, and interaction.

Consequently, current auditory intelligence systems remain fragmented and functionally narrow, lacking the integrative structure characteristic of human auditory cognition. They excel at predefined benchmarks but struggle to generalize across situations or to explain their predictions in meaningful ways. Advancing beyond these constraints requires a paradigm that treats auditory intelligence not as a set of isolated tasks, but as a situated, layered, and explainable process. The next section introduces such paradigms, each targeting a distinct cognitive layer of auditory understanding.

\section{Proposed Paradigms}
\begin{table*}[th]
\caption{Summary of task formulations with outputs, evaluation sketches, and primary uses.}
\centering
\small
\setlength{\tabcolsep}{3pt}
\begin{tabularx}{\linewidth}{l l l l}
\toprule
Task & Output & Eval (sketch) & Primary use \\
\midrule
ASPIRE &
spectro-temporal pattern description &
evidence alignment, coverage, agreement &
explainable recognition \\
SODA &
hierarchical description &
structure fidelity, retrieval, consistency &
perception, retrieval \\
AUX &
causal/explanatory sentences &
cause--effect QA, counterfactual consistency, faithfulness &
diagnostics, decision support \\
AUGMENT &
trigger, event, motivation, goal &
slot F1, inter-annotator agreement, counterfactual stress &
intent-aware interaction \\
\bottomrule
\end{tabularx}
\label{tab:tasks}
\end{table*}

While I have worked on SED and SER to improve accurate recognition of sound events, methods focused on mere detection and labeling now feel mature and yield diminishing returns \cite{mytechreport, filtaug, FDY, coughcam, etri, dcase2024mytechrep, DFD, PFD, TFD, myphddissertation}. It is time to challenge beyond what was heard. To move toward cognitively grounded auditory intelligence, I propose a set of paradigm-level tasks aligned with the layered structure introduced in Section 2. Each formulation targets a distinct facet of human-like auditory cognition—event interpretation, causal explanation, intent/goal inference, and structure-based evidence for explainable recognition. These paradigms are not meant to replace existing tasks; rather, they extend them by reflecting how sound functions in real-world settings and by providing a scaffold that links recognition, reasoning, and interaction.

\subsection{Acoustic Structure-based Parsing and Interpretation for REcognition (ASPIRE)}
ASPIRE is a foundational task that supplies structured evidence to the other three paradigms by enabling detailed, reasoned recognition of sound. It starts from a simple question: spectrograms render sound into spectro-temporal patterns that trained listeners can partially read. Can we train an acoustic model (AM) to parse these patterns into a textual or symbolic description, and a language model (LM) to project that description into higher-level, human-readable sound narratives?

Concretely, ASPIRE aims to textualize spectrograms into pattern descriptions such as: “broadband noise from 1.5--4kHz for 0.8s, followed by a soft harmonic tone centered at 200Hz lasting 2s.” The AM produces low-level spectro-temporal descriptors (e.g., bands, harmonics, transients, onsets/offsets, durations, frequency ranges), and the LM uses this description to generate grounded sound descriptions or to support downstream reasoning.

Although ASPIRE introduces an intermediate representation between waveform and labels, it yields three benefits: (i) explainability—predictions can be traced to explicit spectro-temporal evidence; (ii) granularity—models can reference fine acoustic attributes beyond coarse event tags; and (iii) transferability—the same pattern vocabulary can support diverse tasks (e.g., SODA, AUX, AUGMENT). In practice, such structured descriptions surface subtle cues—early indicators of machine faults, respiratory abnormalities in coughs, or affective prosody in speech—thereby improving both the credibility and utility of audio understanding. Moreover, this representation enables more specific recognition and attribution: models can localize which machine parts likely signal a fault, use cough patterns as screening evidence for candidate conditions, or infer speaker identity and affect with explicit, inspectable support.

I propose to initialize ASPIRE with single-event descriptions drawn from a limited but diverse set of sound events (e.g., tonal, impulsive, broadband, harmonic, modulated) that cover the major spectro-temporal primitives. To bootstrap, the DESED foreground set can be used, although these excerpts are not perfectly isolated and may contain residual background or annotation noise \cite{DCASEtask4}. Expert annotators would provide concise textual spectro-temporal descriptions, with a normalized vocabulary of primitives (e.g., band-limited noise, harmonic stacks, onsets/offsets, AM/FM) and quality control via double annotation and adjudication. Because many acoustic events are compositional, LLMs can map these pattern descriptions to event concepts, supporting generalization beyond the seed classes. After this stage, the learned ASPIRE model can be extended to multi-event (polyphonic) scenes, leveraging compositionality to parse overlapping events and richer contexts.

\subsection{Structured Open Description of Acoustics (SODA)}
SODA generates structured, multi-level descriptions of acoustic scenes by decomposing audio into event, context, and scene components. Unlike SED, ASC, or AAC—which treat sound as discrete events, global labels, or free-form captions—SODA adopts a hierarchical representation aligned with human perception: first identifying atomic events, then grouping them into situational contexts, and finally inferring the encompassing scene. The framework is open-vocabulary and open-structure, allowing outputs to be expressed as natural language with an embedded hierarchy or as a JSON-like schema. This enables flexible, extensible representations of real-world soundscapes and provides a common scaffold for retrieval, explainability, and downstream reasoning.

While recent work has explored open-vocabulary SED \cite{flam}, I extend this notion to open description across the full hierarchy of events, contexts, and scenes. In SODA, a structured multi-task formulation links these levels so that outputs from earlier stages inform later ones: event hypotheses condition contextual descriptions, which in turn support scene inference (either cascaded or jointly trained).

ALMs such as CLAP enable caption-conditioned, open-set recognition by aligning acoustic and textual embeddings. Leveraging ALMs allows SED/ASC—and SODA—to express predictions as captions or structured descriptions and to generalize to unseen classes via text prompts. Moreover, by aligning the AM, or ALM, and (optionally) ASPIRE’s spectro-temporal descriptors with an LLM for lightweight reasoning, the system can recognize events and scenes defined only by textual prototypes or definitions, reducing reliance on class-specific labels and expanding beyond the original training set.

To build SODA, I propose the following steps:
\begin{enumerate}
\item Dataset curation: assemble a corpus jointly labeled for SED, AAC, and ASC. For this purpose I am developing the Hierarchical Extensive Acoustic Recognition (HEAR) dataset from DESED foreground set \cite{DCASEtask4}.
\item Hierarchy-aware training: validate that hierarchy-based learning (cascaded or jointly trained event→context→scene) outperforms single-task and flat multi-task baselines. In this phase, SED and ASC may remain closed-set.
\item Open-description extension: augment the dataset with caption-style annotations and text prototypes to enable open-description outputs; define a JSON-like schema that encodes event, context, and scene.
\item Evaluation protocol: measure structure fidelity (event→context→scene consistency), recognition performance on seen classes, generalization to unseen classes via text prompts, and data efficiency.
\end{enumerate}

SODA enables AMs and ALMs to recognize and describe acoustic environments more generally and robustly by (i) expanding recognition across a three-level hierarchy (event→context→scene) and (ii) shifting from closed-set labels to open-vocabulary, caption-style outputs. I view SODA as a natural next step beyond SED, AAC, and ASC—a unifying formulation that preserves their strengths while addressing their limitations—and as a path for the DCASE community to evolve toward more flexible, explainable, and generalizable benchmarks.

\subsection{Acoustic Understanding and eXplanation (AUX)}
AUX targets the generation of causal and explanatory descriptions for acoustic events. In contrast to AAC, which answers what was heard, AUX focuses on why it happened—inferring plausible causes, likely consequences, and emotional or intentional subtext. This requires integrating background knowledge and performing reasoning beyond the observed audio, linking acoustic evidence to situational context and commonsense. By aligning low-level perception with high-level narrative inference, AUX provides structured rationales (e.g., “because… therefore…”) that can support decision making in safety monitoring, diagnostics, and human–AI interaction.

Current AAC relies on datasets such as Clotho, AudioCaps, and WavCaps \cite{clotho, audiocaps, wavcaps}. Clotho captions were collected via crowdsourcing with an explicit “describe only what is heard” guideline, whereas the latter two involve captions generated or expanded with LLMs, which often introduce implicit inferences beyond audible evidence. In practice, even Clotho contains instances of inferred content in its metadata. I therefore propose to factor captions into two components—observation (heard) and explanation (inferred)—and to run ablations that (i) train and evaluate models on observation-only targets to measure audio grounding, then (ii) add explanation targets to quantify reasoning benefits and any leakage from textual priors.

I propose re-annotating Clotho into two parallel tracks—observations and explanations—and prompting an LLM to produce five items of each per audio clip under strict guidelines: observations must be strictly grounded in audible evidence with no inference, while explanations must provide causal or contextual hypotheses rather than restating what is heard. Using this split, I propose to compare (i) observation-only training and evaluation (true AAC), (ii) explanation-only training (AUX), (iii) mixed multitask learning, and (iv) two-stage learning that first learns AAC and then fine-tunes for AUX; the two-stage setup is expected to benefit AUX by leveraging the provided observations. In addition, I will condition AUX on HEAR outputs from SODA (e.g., event timing, duration, and loudness) so that explanations can reference explicit spectro-temporal evidence when inferring scene context.

By developing AUX, I aim to push AAC beyond observation into sound-based inference about the physical and social world. Concretely, SODA-style acoustic models can produce structured recognition outputs (e.g., events with timing, loudness, and contextual cues), which are then passed to a language model to infer plausible causes, consequences, and intent. ASPIRE complements this pipeline by supplying fine-grained spectro-temporal evidence that grounds the explanations in observable patterns. Together, SODA→AUX (with ASPIRE) enables models to move from “what was heard” to “why it happened” and “what is likely next,” improving causal robustness, interpretability, and generalization beyond closed label sets.

\subsection{Acoustic Understanding via Goal, Motivation, Event, and Trigger (AUGMENT)}
AUGMENT introduces a semantic slot-filling framework that extracts the structural meaning behind sound: who or what initiated it (trigger), what happened (event), why it happened (motivation), and what outcome was intended (goal). Drawing on cognitive theories of narrative and intentionality, AUGMENT reframes auditory understanding as goal-driven inference, linking perceived events to hypothesized agents, motives, and ends. The output is a four-slot record {trigger, event, motivation, goal} expressed in natural language or symbols, optionally accompanied by confidence scores and pointers to evidence (e.g., timestamps, ASPIRE descriptors, SODA context). This structure supports narrative reconstruction, intent-aware human–robot interaction, and safety diagnostics, and can be evaluated via slot-level accuracy/F1, inter-annotator agreement, and robustness to counterfactual perturbations.

Building on AUX, AUGMENT decomposes sound-related context into more specific components. Not every recording will instantiate all four slots (trigger, event, motivation, goal), but the schema yields a deeper, more structured account when applicable. When acoustic models and audio–language models can reliably perform AUGMENT, much of the information we typically infer from sound—agents, actions, reasons, and intended outcomes—can be extracted directly from audio-driven evidence. This represents a substantive step toward comprehensive auditory intelligence, linking low-level cues to actionable, intent-aware understanding.

\subsection{Generative Auditory Intelligence}
Recent DCASE challenges have explored label-to-audio and text-to-audio generation for synthesizing sound effects applicable to film, games, and XR, enabling more realistic and responsive experiences. I argue that the next step is to couple generation tightly with perception and reasoning: conditioning sound synthesis on structured outputs from ASPIRE (spectro-temporal descriptors), SODA (event–context–scene), and AUX/AUGMENT (causes, intent, goals). This linkage yields controllable, explainable, and scene-consistent audio—e.g., foley that matches inferred actions, speech whose style aligns with context and affect, and music that reflects situational goals.

In practice, text-to-audio should be integrated with text-to-speech and text-to-music so that AI assistants and robots can produce not only sound events but also speech and music that are context-aware and environment-dependent. Evaluation should assess perceptual quality (e.g., human ratings), semantic alignment with the conditioning signals, temporal and spatial coherence within the scene, and safety constraints such as appropriate loudness. This generative capability completes the loop from recognition and reasoning to action, enabling agents to communicate and intervene through sound in a principled way.

\subsection{Toward Multimodal Audio-Text-Image Grounding}
These paradigms can be extended to multimodal grounding by linking auditory events with linguistic and visual representations. Aligning concepts across sound, language, and vision enables richer situational awareness, zero-shot generalization to unseen categories, and visually grounded explanations and predictions. For example, a structured caption produced by SODA (optionally enriched with ASPIRE descriptors and AUX/AUGMENT inferences) can be used to retrieve or generate a representative image, constraining visual content to remain consistent with the inferred scene, causes, intent, and goals.

Practically, this can be bootstrapped from existing audio–text corpora by augmenting them with synthetic or retrieved images to form audio–text–image triplets. Clotho’s five captions per clip can be converted into multiple scene renderings, producing diverse visualizations of the same audio. When HEAR is available, its sound-event and acoustic-scene labels can be composed into prompts to generate images that reflect both fine-grained events and global context. The resulting collection yields not only full triplets but also audio–image pairs for cross-modal learning and retrieval.

Evaluation should measure cross-modal retrieval and alignment (e.g., Recall@K), semantic and structural consistency between audio-derived descriptions and images, and the utility of visual grounding for downstream audio tasks (such as improved recognition or explanation quality). Because humans rely heavily on both vision and hearing, the longer-term goal is integrated audio–visual intelligence that endows robots, machines, AI agents, and future devices (e.g., smart glasses) with human-aligned sensory understanding—enabling seamless coordination with human perception and augmenting our sensing across time and environments.

\section{Discussion}
The proposed paradigms aim to bridge the gap between low-level auditory perception and high-level cognitive reasoning, situating sound within a broader framework of understanding. Rather than treating sound solely as a classification target or descriptive input, they construe auditory signals as cognitive evidence—inputs that invite inference, explanation, and interaction.

This reframing mirrors human hearing: we recognize acoustic patterns, infer causes, assess affect and intent, and anticipate likely continuations. These functions are situationally integrated rather than task-specific, enabling efficient navigation of complex environments and social dynamics. Aligning machine hearing with this structure can yield systems that are more generalizable, explainable, and responsive.

The paradigms also advance explainable AI by making audio-based decisions interpretable at multiple levels—what was heard, why it was heard, and what might follow—and they lay groundwork for multimodal understanding in which audio, language, and vision support unified perceptual reasoning. The capacity to predict future situations from sound enables anticipatory intelligence, a key ingredient for robotics, autonomous systems, and real-time decision-making.

Several open challenges remain: curating sufficiently rich datasets; developing reasoning models that generalize; and evaluating explanations and predictions in realistic settings. Bridging the gap between low-level acoustic features and high-level symbolic reasoning will require progress in representation learning and cognitive modeling. Addressing these challenges will benefit from collaboration across machine learning, cognitive science, and auditory neuroscience.

\section{Conclusion}
This paper has argued for a paradigm shift in how auditory intelligence is conceptualized and developed. Moving beyond recognition-based tasks, I proposed a set of cognitively inspired paradigms—ASPIRE, SODA, AUX, and AUGMENT—that reframe machine hearing as a layered process of perception, reasoning, and interaction. These paradigms are not isolated benchmarks, but structural scaffolds for developing more human-aligned, context-aware auditory intelligence. By grounding sound in its situational, emotional, and intentional dimensions, they align machine hearing more closely with human auditory cognition, supporting not only more explainable and generalizable models but also new applications in robotics, assistive technology, and autonomous systems. The path forward will require cross-disciplinary collaboration and the development of new tools, datasets, and evaluation methods—but the goal is clear: to build systems that do not merely hear, but understand.

I offer this work as both a proposal and an invitation—for researchers, designers, and practitioners—to reconsider the role of sound in artificial intelligence and to explore what it truly means for machines to listen as a form of intelligence.

\bibliographystyle{IEEEtran}
\bibliography{refs}

\begin{thebibliography}{10}
\providecommand{\url}[1]{#1}
\csname url@samestyle\endcsname
\providecommand{\newblock}{\relax}
\providecommand{\bibinfo}[2]{#2}
\providecommand{\BIBentrySTDinterwordspacing}{\spaceskip=0pt\relax}
\providecommand{\BIBentryALTinterwordstretchfactor}{4}
\providecommand{\BIBentryALTinterwordspacing}{\spaceskip=\fontdimen2\font plus
\BIBentryALTinterwordstretchfactor\fontdimen3\font minus \fontdimen4\font\relax}
\providecommand{\BIBforeignlanguage}[2]{{%
\expandafter\ifx\csname l@#1\endcsname\relax
\typeout{** WARNING: IEEEtran.bst: No hyphenation pattern has been}%
\typeout{** loaded for the language `#1'. Using the pattern for}%
\typeout{** the default language instead.}%
\else
\language=\csname l@#1\endcsname
\fi
#2}}
\providecommand{\BIBdecl}{\relax}
\BIBdecl

\bibitem{GPT4}
OpenAI, ``Gpt-4 technical report,'' \emph{arXiv preprint arXiv:2303.08774}, 2024.

\bibitem{llama2}
MetaGenAI, ``Llama 2: Open foundation and fine-tuned chat models,'' \emph{arXiv preprint arXiv:2307.09288}, 2023.

\bibitem{deepseek}
DeepSeek-AI, ``Deepseek-v3 technical report,'' \emph{arXiv preprint arXiv:2412.19437}, 2025.

\bibitem{CASSE}
T.~Virtanen, M.~D. Plumbley, and D.~Ellis, \emph{Computational Analysis of Sound Scenes and Events}, 1st~ed.\hskip 1em plus 0.5em minus 0.4em\relax Springer Publishing Company, Incorporated, 2017, pp. 3--11, 71--77.

\bibitem{DCASEtask4}
N.~Turpault, R.~Serizel, A.~P.~Shah, and J.~Salamon, ``{Sound event detection in domestic environments with weakly labeled data and soundscape synthesis},'' in \emph{{DCASE Workshop}}, 2019.

\bibitem{crnn}
E.~{\c{C}}ak{\i}r, G.~Parascandolo, T.~Heittola, H.~Huttunen, and T.~Virtanen, ``Convolutional recurrent neural networks for polyphonic sound event detection,'' \emph{IEEE/ACM Transactions on Audio, Speech, and Language Processing}, vol.~25, no.~6, pp. 1291--1303, 2017.

\bibitem{sedmetrics}
A.~Mesaros, T.~Heittola, and T.~Virtanen, ``Metrics for polyphonic sound event detection,'' \emph{Applied Sciences}, vol.~6, no.~6, 2016.

\bibitem{PSDS}
{\c{C}}.~Bilen, G.~Ferroni, F.~Tuveri, J.~Azcarreta, and S.~Krstulovi\'{c}, ``A framework for the robust evaluation of sound event detection,'' in \emph{ICASSP}, 2020, pp. 61--65.

\bibitem{freqdeptalsp}
H.~Nam, S.-H. Kim, D.~Min, B.-Y. Ko, and Y.-H. Park, ``Towards understanding of frequency dependence on sound event detection,'' \emph{arXiv preprint arXiv:2502.07208}, 2025.

\bibitem{jitter}
H.~Nam and Y.-H. Park, ``Jitter: Jigsaw temporal transformer for event reconstruction for self-supervised sound event detection,'' \emph{arXiv preprint arXiv:2502.20857}, 2025.

\bibitem{specaug}
D.~S. Park, W.~Chan, Y.~Zhang, C.-C. Chiu, B.~Zoph, E.~D. Cubuk, and Q.~V. Le, ``{SpecAugment: A Simple Data Augmentation Method for Automatic Speech Recognition},'' in \emph{Proc. Interspeech}, 2019.

\bibitem{conformer}
A.~Gulati, J.~Qin, C.-C. Chiu, N.~Parmar, Y.~Zhang, J.~Yu, W.~Han, S.~Wang, Z.~Zhang, Y.~Wu, and R.~Pang, ``{Conformer: Convolution-augmented Transformer for Speech Recognition},'' in \emph{Proc. Interspeech}, 2020.

\bibitem{mpc}
H.~Nam and Y.-H. Park, ``Coherence-based phonemic analysis on the effect of reverberation to practical automatic speech recognition,'' \emph{Applied Acoustics}, vol. 227, p. 110233, 2025.

\bibitem{wav2vec2.0}
A.~Baevski, Y.~Zhou, A.~Mohamed, and M.~Auli, ``wav2vec 2.0: A framework for self-supervised learning of speech representations,'' in \emph{Advances in Neural Information Processing Systems}, 2020.

\bibitem{hubert}
W.-N. Hsu, B.~Bolte, Y.-H.~H. Tsai, K.~Lakhotia, R.~Salakhutdinov, and A.~Mohamed, ``Hubert: Self-supervised speech representation learning by masked prediction of hidden units,'' \emph{IEEE/ACM Transactions on Audio, Speech, and Language Processing}, 2021.

\bibitem{ASP}
K.~Okabe, T.~Koshinaka, and K.~Shinoda, ``Attentive statistics pooling for deep speaker embedding,'' in \emph{Proc. Interspeech}, 2018.

\bibitem{SAP}
W.~Cai, J.~Chen, and M.~Li, ``Exploring the encoding layer and loss function in end-to-end speaker and language recognition system,'' in \emph{Proc. Interspeech}, 2018.

\bibitem{tdyaccess}
S.-H. Kim, H.~Nam, and Y.-H. Park, ``Analysis-based optimization of temporal dynamic convolutional neural network for text-independent speaker verification,'' \emph{IEEE Access}, vol.~11, 2023.

\bibitem{freqse}
J.~Thienpondt, B.~Desplanques, and K.~Demuynck, ``Integrating frequency translational invariance in tdnns and frequency positional information in 2d resnets to enhance speaker verification,'' in \emph{Proc. Interspeech}, 2021.

\bibitem{c2datt}
J.~Li, Y.~Tian, and T.~Lee, ``Convolution-based channel-frequency attention for text-independent speaker verification,'' in \emph{ICASSP}, 2023.

\bibitem{PANN}
Q.~Kong, Y.~Cao, T.~Iqbal, Y.~Wang, W.~Wang, and M.~D. Plumbley, ``Panns: Large-scale pretrained audio neural networks for audio pattern recognition,'' \emph{IEEE/ACM Transactions on Audio, Speech, and Language Processing}, 2020.

\bibitem{coughcam}
G.-T. Lee, H.~Nam, S.-H. Kim, S.-M. Choi, Y.~Kim, and Y.-H. Park, ``Deep learning based cough detection camera using enhanced features,'' \emph{Expert Systems with Applications}, vol. 206, 2022.

\bibitem{etri}
S.-H. Kim, H.~Nam, S.-M. Choi, and Y.-H. Park, ``Real-time sound recognition system for human care robot considering custom sound events,'' \emph{IEEE Access}, vol.~12, 2024.

\bibitem{AST}
Y.~Gong, Y.-A. Chung, and J.~Glass, ``Ast: Audio spectrogram transformer,'' in \emph{Proc. Interspeech}, 2021.

\bibitem{beats}
S.~Chen, Y.~Wu, C.~Wang, S.~Liu, D.~Tompkins, Z.~Chen, W.~Che, X.~Yu, and F.~Wei, ``Beats: Audio pre-training with acoustic tokenizers,'' in \emph{ICML}, 2023.

\bibitem{seld2019}
A.~Politis, A.~Mesaros, S.~Adavanne, T.~Heittola, and T.~Virtanen, ``Overview and evaluation of sound event localization and detection in dcase 2019,'' \emph{IEEE/ACM Transactions on Audio, Speech, and Language Processing}, vol.~29, 2020.

\bibitem{starss22}
A.~Politis, K.~Shimada, P.~Sudarsanam, S.~Adavanne, D.~Krause, Y.~Koyama, N.~Takahashi, S.~Takahashi, Y.~Mitsufuji, and T.~Virtanen, ``{STARSS22}: {A} dataset of spatial recordings of real scenes with spatiotemporal annotations of sound events,'' in \emph{{DCASE Workshop}}, 2022.

\bibitem{2022t3report}
B.-Y. Ko, H.~Nam, S.-H. Kim, D.~Min, S.-D. Choi, and Y.-H. Park, ``Data augmentation and squeeze-and-excitation network on multiple dimension for sound event localization and detection in real scenes,'' DCASE Challenge, Tech. Rep., 2022.

\bibitem{Biseld}
G.-T. Lee, H.~Nam, and Y.-H. Park, ``Binaural sound event localization and detection based on hrtf cues for humanoid robots,'' \emph{arXiv preprint arXiv:2507.20530}, 2025.

\bibitem{dcaseaac}
K.~Drossos, S.~Adavanne, and T.~Virtanen, ``Automated audio captioning with recurrent neural networks,'' in \emph{IEEE Workshop on Applications of Signal Processing to Audio and Acoustics}, 2017.

\bibitem{clotho}
K.~Drossos, S.~Lipping, and T.~Virtanen, ``Clotho: an audio captioning dataset,'' in \emph{ICASSP}, 2020.

\bibitem{chatgptaugaac}
I.~Choi, H.~Nam, D.~Min, S.-D. Choi, and Y.-H. Park, ``Chatgpt caption paraphrasing and fense-based caption filtering for automated audio captioning,'' DCASE Challenge, Tech. Rep., 2024.

\bibitem{dcasebed2024}
J.~Liang, I.~Nolasco, B.~Ghani, H.~Phan, E.~Benetos, and D.~Stowell, ``{Mind the Domain Gap: a Systematic Analysis on Bioacoustic Sound Event Detection},'' \emph{arXiv preprint arXiv:2403.18638}, 2024.

\bibitem{bioacousticstrf}
D.~Min, H.~Nam, and Y.-H. Park, ``Few-shot bioacoustic event detection utilizing spectro-temporal receptive field,'' in \emph{{Proc. INTER-NOISE}}, 2024.

\bibitem{prtfnet}
B.-Y. Ko, G.-T. Lee, H.~Nam, and Y.-H. Park, ``Prtfnet: Hrtf individualization for accurate spectral cues using a compact prtf,'' \emph{IEEE Access}, vol.~11, 2023.

\bibitem{brainstem}
B.-Y. Ko, Y.-H. Park, G.-T. Lee, and H.~Nam, ``Filteraugment: An acoustic environmental data augmentation method,'' in \emph{International Congress on Acoustics (ICA)}, 2022.

\bibitem{contHRTF}
B.-Y. Ko, D.~Min, H.~Nam, and Y.-H. Park, ``Dnn based hrirs identification with a continuously rotating speaker array,'' \emph{arXiv preprint arXiv:2504.14817}, 2025.

\bibitem{audioldm}
H.~Liu, Z.~Chen, Y.~Yuan, X.~Mei, X.~Liu, D.~Mandic, W.~Wang, and M.~D. Plumbley, ``{A}udio{LDM}: Text-to-audio generation with latent diffusion models,'' in \emph{ICML}, 2023.

\bibitem{audiogen}
F.~Kreuk, G.~Synnaeve, A.~Polyak, U.~Singer, A.~D{\'e}fossez, J.~Copet, D.~Parikh, Y.~Taigman, and Y.~Adi, ``Audiogen: Textually guided audio generation,'' in \emph{International Conference on Learning Representations (ICLR)}, 2023.

\bibitem{vifs}
J.~Lee, H.~Nam, and Y.-H. Park, ``Vifs: An end-to-end variational inference for foley sound synthesis,'' DCASE Challenge, Tech. Rep., 2023.

\bibitem{mytechreport}
H.~Nam, B.-Y. Ko, G.-T. Lee, S.-H. Kim, W.-H. Jung, S.-M. Choi, and Y.-H. Park, ``Heavily augmented sound event detection utilizing weak predictions,'' DCASE Challenge, Tech. Rep., 2021.

\bibitem{filtaug}
H.~Nam, S.-H. Kim, and Y.-H. Park, ``Filteraugment: An acoustic environmental data augmentation method,'' in \emph{ICASSP}, 2022.

\bibitem{FDY}
H.~Nam, S.-H. Kim, B.-Y. Ko, and Y.-H. Park, ``{Frequency Dynamic Convolution: Frequency-Adaptive Pattern Recognition for Sound Event Detection},'' in \emph{Proc. Interspeech}, 2022.

\bibitem{dcase2024mytechrep}
H.~Nam, D.~Min, I.~Choi, S.-D. Choi, and Y.-H. Park, ``Self training and ensembling frequency dependent networks with coarse prediction pooling and sound event bounding boxes,'' DCASE Challenge, Tech. Rep., 2024.

\bibitem{DFD}
H.~Nam, S.-H. Kim, D.~Min, J.~Lee, and Y.-H. Park, ``Diversifying and expanding frequency-adaptive convolution kernels for sound event detection,'' in \emph{Proc. Interspeech}, 2024.

\bibitem{PFD}
H.~Nam and Y.-H. Park, ``Pushing the limit of sound event detection with multi-dilated frequency dynamic convolution,'' \emph{arXiv preprint arXiv:2406.13312}, 2024.

\bibitem{TFD}
------, ``Temporal attention pooling for frequency dynamic convolution in sound event detection,'' \emph{arXiv preprint arXiv:2504.12670}, 2025.

\bibitem{myphddissertation}
H.~Nam, ``Frequency dynamic convolutions for sound event detection,'' \emph{arXiv preprint arXiv:2506.12785}, 2025.

\bibitem{flam}
Y.~Wu, C.~Tsirigotis, K.~Chen, C.-Z.~A. Huang, A.~Courville, O.~Nieto, P.~Seetharaman, and J.~Salamon, ``Flam: Frame-wise language-audio modeling,'' \emph{arXiv preprint arXiv:2505.05335}, 2025.

\bibitem{audiocaps}
C.~D. Kim, B.~Kim, H.~Lee, and G.~Kim, ``{A}udio{C}aps: Generating captions for audios in the wild,'' in \emph{Proceedings of the 2019 Conference of the North {A}merican Chapter of the Association for Computational Linguistics: Human Language Technologies, Volume 1 (Long and Short Papers)}, 2019.

\bibitem{wavcaps}
X.~Mei, C.~Meng, H.~Liu, Q.~Kong, T.~Ko, C.~Zhao, M.~D. Plumbley, Y.~Zou, and W.~Wang, ``Wavcaps: A chatgpt-assisted weakly-labelled audio captioning dataset for audio-language multimodal research,'' \emph{IEEE/ACM Transactions on Audio, Speech, and Language Processing}, 2024.

\end{thebibliography}

\end{document}